\documentclass[12pt]{article}

\makeatletter
\@addtoreset{equation}{section}
\makeatother

\def\be{\begin{equation}}
\def\ee{\end{equation}}
\newcommand{\bee}{\begin{eqnarray}}
\newcommand{\eee}{\end{eqnarray}}
\newcommand{\nn}{\nonumber}

\newcommand{\go}{\omega}
\newcommand{\f}{\frac}
\newcommand{\q}{\,,\qquad}

\newcommand{\p}{\partial}
\newcommand{\e}{\epsilon}
\newcommand{\half}{\frac{1}{2}}
\newcommand{\ga}{\alpha}

\newcommand{\M}{{\cal M}}

\newcommand{\dga}{{\dot{\alpha}}}
\newcommand{\dgb}{{\dot{\beta}}}
\newcommand{\gb}{\beta}
\newcommand{\gga}{\gamma}

\newcommand{\gvep}{\varepsilon}
\newcommand{\gs}{\sigma}

\newcommand{\R}{{\cal R}}
\newcommand{\W}{{\cal W}}
\newcommand{\C}{{\cal C}}

\newcommand{\gee}{\epsilon}

\newcommand{\D}{{\cal D}}

\begin{document}

\begin{center}
{\large\bf{Higher-Spin Theories and $Sp(2M)$ Invariant Space--Time}}
\vglue 0.6  true cm
\vskip1cm
\vskip0.5cm
M.A.~Vasiliev
\vglue 0.3  true cm

I.E.Tamm Department of Theoretical Physics, Lebedev Physical Institute,\\
Leninsky prospect 53, 119991, Moscow, Russia
\vskip1.5cm
\end{center}



\begin{abstract}
Some methods of the ``unfolded dynamics'' machinery particularly useful
for the analysis of higher spin gauge theories are summarized.
A formulation
of $4d$ conformal  higher spin theories in $Sp(8)$ invariant space-time
with matrix coordinates  and its extension to $Sp(2M)$ invariant
space-times are discussed.  A new result on the global characterizaton
of  causality of physical events in the $Sp(2M)$ invariant space-time
is announced.
\end{abstract}


\section{Introduction}

Higher spin (HS) gauge theories are gauge theories with tensorial gauge
symmetry parameters. In the Fronsdal`s formulation~\cite{Fron},
for a spin $s$  totally
symmetric gauge field $\phi_{n_1 \ldots n_{s}}$,
the gauge parameter $\gee_{n_1 \ldots n_{s-1}}$
 is totally symmetric in its indices and
traceless $\gee^m{}_{m n_3 \ldots n_{s-1}}=0$. The field
$\phi_{n_1 \ldots n_{s}}$ is double traceless
$\phi^{kl}{}_{kl n_5 \ldots n_{s}}=0$. (For other
equivalent approaches to HS massless field see, e.g.,~\cite{diff}).
For every spin $s$ there
exists~\cite{Fron} a unique action with at most
two derivatives
$
S=\int d^d x \partial_{\ldots} \phi_{\ldots} (x)\partial_{\ldots}
\phi_{\ldots} (x)
$
invariant under the Abelian gauge transformations
\be
\delta \phi_{n_1\ldots n_s} (x)= \partial_{\{n_1}\gee_{n_2 \ldots n_{s}\}}(x)
,
\qquad \partial_n = \frac{\partial}{ \partial x^n}.
\ee
For $s=1$ and 2 this reproduces the Maxwell theory and linearized
Einstein theory. Fields of half-integer spins are  described
analogously. Since the lower spin gauge theories with spins
$s=1,3/2$ and 2 play a key role in the theory of fundamental
interactions it is  interesting to understand whether there is
some nontrivial theory behind spin $s>2$ gauge fields,
the HS gauge fields.

There is a number of motivations for studying HS gauge theories.
{}From supergravity perspective, this is interesting because theories
with HS fields may have more supersymmetries than the ``maximal''
supergravities with 32 supercharges like $11d$ SUGRA~\cite{CJS}.
Recall that the limitation that the number of supercharges is $\leq
32$ is a direct consequence of the requirement that $s\leq 2$ for all
fields in a supermultiplet (see, e.g.,~\cite{PVN}).  {}From
superstring perspective, most obvious motivation is due to
Stueckelberg symmetries in the string field theory~\cite{SFT}, which
have a form of some spontaneously broken HS gauge symmetries.
Whatever a symmetric phase of the superstring theory is, Stueckelberg
symmetries are expected to become unbroken HS symmetries in such a
phase and, therefore, the superstring field theory has to become one
or another version of the HS gauge theory.  An important indication in
the same direction is ~\cite{Gross} that string amplitudes exhibit
certain symmetries in the high-energy limit equivalent to the string
mass parameter tending to zero.  In all cases, the key issue is the HS
gauge symmetry.

Unusual feature of the interacting HS gauge theory is that unbroken HS
gauge symmetries do not allow flat space-time as a vacuum solution,
requiring nonzero space-time curvature with anti-de Sitter (AdS)
space-time as a most symmetric vacuum~\cite{FV1}.  Recently, this
unusual property received
interpretation~\cite{Sund,Wit2,SS1,BHS,AAT,KP} in the context of the
AdS/CFT correspondence conjecture~\cite{JM}. In particular,
in~\cite{Sund,Wit2,KP} it was conjectured that HS gauge theories in
AdS bulk are duals of some conformal models on the AdS boundary in the
large $N$ limit with $g^2 N \to 0$ where $g^2$ is the boundary
coupling constant, which is opposite to the limit $g^2 N \to \infty $
studied within the original AdS/CFT correspondence
conjecture~\cite{JM}.  Again, this suggests that the HS gauge theory
has a good chance to be related to a symmetric phase of superstring
theory.  On the other hand an explanation why the HS gauge theory has
not been yet observed in the superstring theory may be just that no
complete formulation of the latter is still known in the AdS
background at the quantum level despite the progress achieved at the
classical level~\cite{MT}.

Properties of the HS theories can to large extent be revealed from the
structure of global HS symmetry algebras found originally
in~\cite{FVa} for the simplest HS model in $AdS_4$. In~\cite{Fort2} it
was realized that $AdS_4$ HS algebras are certain star-product
algebras with spinorial generating elements, while in~\cite{VF} it was
then conjectured that algebras of the same structure correspond to HS
models in higher dimensions.\footnote{Recently, analogous realization
  of the string field theory star product in terms of infinite set of
  oscillators carrying vector space-time indices was proposed
  in~\cite{Bars}.}  The construction is as follows. Consider
oscillators $a_A$ and $b^B$ satisfying the commutation relations
\be
[a_A, b^B ] = \delta_A^B.
\ee
In HS applications, indices $A,B= 1,2\ldots M$ are interpreted as
spinorial (then $M= 2^p$). The HS fields are described by the gauge
potentials
\be
\label{expgo}
dx^n \go_n (a,b|x) =
\sum_{p,q} dx^n \go_{n A_1 \ldots A_p}{}^{B_1 \ldots B_q} (x)
b^{A_1} \ldots b^{A_p} a_{B_1} \ldots a_{B_q}\,
\ee
taking values in the oscillator algebra.  The field $dx^n \go_n
(a,b|x)$ is the HS generalization of the frame field and Lorentz
connection in the Cartan formulation of gravity. The HS curvature and
transformation law have the standard Yang--Mills form with the
oscillator algebra in place of a Yang--Mills matrix algebra
\be
R= d\go +\go\wedge \go,\qquad d = dx^n \p_n,
\ee
\be
\delta \go   = d\gvep +
[\go,  \gvep ].
\ee
The HS gauge symmetry parameter $\gvep(a,b |x)$ is an arbitrary
function of the space-time coordinates $x^n$ and noncommutative
auxiliary variables $ a_A$ and $b^B$, which admits an expansion
analogous to (\ref{expgo}). Since this expansion contains
infinitely many terms,       the HS algebra of oscillators is
infinite-dimensional. It contains however finite-dimensional
subalgebras. Important example is given by the subalgebra $sp(2M)$
spanned by various bilinears in the oscillators
\be
\label{sp}
P_{AB} = a_A a_B ,\qquad L_A{}^B = \half \{a_A,  b^B\}
,\qquad K^{AB} = b^A b^B \,
\ee
and its superextension $osp(1,2M)$ with supercharges $Q_A= a_A$ and
$S^B= b^B$.  Usual $AdS$ and conformal symmetry algebras belong to
this $sp(2M)$. In particular, $AdS_3$ algebra $o(2,2)$ is isomorphic
to $sp(2)\oplus sp(2)$. (The doubling which extends to the whole
$AdS_3$ HS system~\cite{Blen} is introduced with the aid of additional
involutive element $\psi$ ($\psi^2 =1$)~\cite{Prok}.) The $3d$
conformal (equivalently $AdS_4$) algebra $o(3,2)$ is isomorphic to
$sp(4|R)$. The $4d$ conformal (equivalently $AdS_5$) algebra $o(4,2)$
is isomorphic to $su(2;2)\subset sp(8)$. For higher dimensions usual
space-time symmetry algebras are subalgebras of appropriate $sp(2^p)$.
As was shown long ago in~\cite{HP} $osp(1,2^p)$ provides a natural
simple superextension of the higher dimensional space-time symmetries.

The following important properties of the HS theories follow directly
from the structure of HS algebras.

Usual spin 2 gravitational fields take values in the algebra $sp(2M)$
spanned by bilinears of the oscillators.  Higher spins are associated
with the higher order polynomials of the oscillators.  Because
higher-order polynomials do not close to a finite-dimensional algebra,
it follows that an infinite tower of HS fields should be included once
there is at least one HS field.

By virtue of the field equations, the quantum-mechanical nonlocality
of the HS algebra is translated to the space-time nonlocality of HS
interactions~\cite{Gol}.  Since higher spins require higher
derivatives~\cite{LC,cov}, the theory with infinite towers of higher
spins requires infinitely many derivatives.

More recent observation~\cite{BHS,Mar} we focus on in this report is
that, starting from $d=4$, usual space-time symmetries in conformal HS
models extend to larger symplectic symmetries acting on the conformal
HS multiplets. We will see that this may affect considerably the
geometric picture of the world leading to new geometries with the so
called ``central charge'' coordinates included. This, however, does
not affect visualization of our world as a four-dimensional space-time
realized as a three-brane embedded into the ten-dimensional
generalized space-time with matrix coordinates.

\section{$4d$ conformal free fields}
\label{$4d$ conformal free fields}
Consider a Fock vacuum state $|0\rangle$ satisfying
\be
a_A |0\rangle = 0 \qquad
\ee
and a set of functions taking values in the Fock space
(i.e., sections of the Fock fiber bundle over space-time)
\be
|\Phi (x) \rangle = C(b|x) |0\rangle,\qquad C(b|x)=\sum_{n=0}^{\infty}
\frac{1}{n!}
C_{A_1 \ldots A_n}(x) b^{A_1} \ldots b^{A_n} .
\ee
To make contact with the description of $4d$ relativistic fields
in terms of two-component indices one replaces the index $A=1\ldots 4$
by a pair of two-component indices $A=(\ga,\dga)$,
($\ga,\gb\ldots  = 1,2$; $\dga,\dgb\ldots  = 1,2$)
and uses the equivalent expansion
\be
\label{2kexp}
C(b,\bar{b}|x)=\sum_{m,n=0}^{\infty}\frac{1}{m!n!}
C_{\ga_1 \ldots \ga_m,\dgb_1 \ldots \dgb_n }(x) b^{\ga_1} \ldots b^{\ga_m}
\bar{b}^{\dgb_1} \ldots \bar{b}^{\dgb_n}.
\ee

The key observation is that the set of all massless conformal
field equations in four dimensions can be formulated in the
following compact form~\cite{BHS}
\be
\label{meq}
D_0 |\Phi (x) \rangle \equiv d |\Phi (x) \rangle + \go_0
|\Phi (x) \rangle =0,\qquad \go_0 = - dx^{\ga\dgb} a_\ga \bar{a}_\dgb.
\ee
Here  $dx^{\ga\dgb}$ is the $4d$ Minkowski  frame 1-form and
$\go_0$ is  the gauge field associated with the
generators of translations realized as bilinear combinations of
mutually commuting oscillators. Note that analogous description of
$3d$ conformal fields was given in~\cite{ShV} (see also \cite{Sh}).

Equation~(\ref{meq}) is equivalent to the equation
\be
\label{oldeq}
\frac{\p}{\p x^{\ga\dgb}} C(b,\bar{b}|x) - \frac{\p^2}{\p b^\ga \p \bar{b}^\dgb}
C(b,\bar{b}|x) =0,
\ee
which decomposes into independent subsystems associated
with different spins, singled out by the condition
\be
\Big (b^\ga\f{\p}{\p b^\ga} - \bar{b}^\dgb \f{\p}{\p \bar{b}^\dgb}\Big)
C(b,\bar{b}|x) = \pm 2s C(b,\bar{b}|x),
\ee
equivalent to
\be
\label{ncond}
N |\Phi (x) \rangle =\pm 2s |\Phi (x) \rangle
,\qquad N= a_\ga b^\ga - \bar{a}_\dgb \bar{b}^\dgb.
\ee

The fact that $4d$ massless field equations can be reformulated in the
form (\ref{oldeq}) was known long ago~\cite{unf,Eis}.  The dynamical
fields are those in the expansion (\ref{2kexp}), carrying either only
dotted or only undotted indices. They are contained in the analytic
($C(b,0|x))$ and antianalytic ($C(0,\bar{b} |x)$) parts, and describe
scalar $C$, spinor $C_\ga b^\ga+ \bar{C}_\dgb \bar{b}^\dgb$, spin 1
field strength $C_{\ga\gb}b^\ga b^\gb +\bar{C}_{\dga\dgb}\bar{b}^\dga
\bar{b}^\dgb$ and so on for spins 3/2 and higher.  All components in
$C(b,\bar{b}|x)$ which depend both on $b$ and on $\bar{b}$ are
auxiliary being expressed by (\ref{oldeq}) in terms of space-time
derivatives of the dynamical fields. The nontrivial equations on the
dynamical fields are
\be
\label{masseq}
\Big (\f{\p^2}{\p b^\gga \p x^{\ga\dgb}}-
\f{\p^2}{\p b^\ga \p x^{\gga\dgb}} \Big) C(b,0|x) =0,\qquad
\Big (\f{\p^2}{\p \bar{b}^\dga \p x^{\ga\dgb}}-
\f{\p^2}{\p \bar{b}^\dgb \p x^{\ga\dga}} \Big) C(0,\bar{b}|x) =0
\ee
for $s>0$ and the Klein--Gordon equation for the spin zero scalar
field $C(0,0|x)$. These are the usual massless equations formulated in
terms of field strengths.  The simple observation~\cite{BHS} that
massless equations admit the described Fock space realization allows
us, with the help of the unfolded dynamics machinery, to reveal their
symmetries which turn out to be much richer than the usual conformal
symmetry of massless fields and, in particular, contain $sp(8)$
symmetry.

\section{Unfolded dynamics}
\label{Unfolded dynamics}

Let $W^a (x)$ be some set of differential $p-$forms with $p\geq 0$
(0-forms are included), and the generalized curvatures $\R^a$ be
defined by the relations
\be
\R^a= dW^a +F^a (W),
\ee
where $F^a$ are some functions of $W^b$ built with the aid of the
exterior product of differential forms.  Given
$F^a (W)$ with vanishing 0-form part ($deg(F^a)>0$) which
satisfies the generalized Jacobi identity
\be
\label{BI}
F^b \frac{\delta F^a }{\delta W^b} \equiv 0,
\ee
we say following~\cite{FDA} that it defines a free differential
algebra. This property guarantees that the generalized Bianchi
identity $ d\R^a = \R^b \frac{\delta F^a }{\delta W^b} $ is satisfied,
which implies that the differential equations on $W^a$ \be
\label{eq}
\R^a =0
\ee
are consistent. These equations are invariant under
the gauge transformations
\be
\label{delw}
\delta W^a = d \e^a -\e^b
\frac{\delta F^a }{\delta W^b},
\ee
where $\epsilon^a (x) $ is an arbitrary $(deg (W^a)-1)- $form (0-forms
do not give rise to gauge parameters) because the generalized
curvatures transform as $ \delta \R^a =-\R^c \frac{\delta }{\delta
  W^c} \left (\e^b \frac{\delta F^a }{\delta W^b} \right).  $ Also,
since Eqs.~(\ref{eq}) are formulated in terms of differential forms,
they are manifestly general coordinate invariant.  Elementary but
crucial fact is that any dynamical system can be ``unfolded'' to the
form (\ref{eq}) by introducing enough auxiliary fields. Note that in
such approach, the co-frame field $e$ from which the metric tensor is
built is one of the 1-forms in the set $W^a$. It is supposed to be
invertible, having a part of order 1 in any perturbative expansion.
This provides a meaningful linearization of Eq.~(\ref{eq}).

For the particular case when the set $W^a$ consists of only 1-forms
$\go^i$, the function $F^i (\go)$ is bilinear $ F^i = f^i_{jk} \go^j
\wedge \go^k $ and the relation (\ref{BI}) amounts to the Jacobi
identity for a Lie algebra $g$ with the structure coefficients $
f^i_{jk} $ (or superalgebra if some of $\go^i$ carry an additional
Grassmann grading).  If the set $W^a$ also contains some $p$-forms
$v^\ga$ (e.g., 0-forms) and the functions $F^\ga$ are linear in $v$, $
F^\ga = t_i{}^\ga {}_\gb \go^i \wedge v^\gb, $ the relation (\ref{BI})
implies that the matrices $t_i{}^\ga {}_\gb$ form some representation
$t$ of $g$ while (\ref{eq}) contains zero-curvature equations of $g$
\be
\label{R0}
R^i =d\go^i +  f^i_{jk} \go^j \wedge \go^k =0
\ee
along with the covariant constancy equation
\be
Dv^\ga=dv^\ga + t_i{}^\ga {}_\gb \go^i \wedge v^\gb =0
\ee
for the representation $t$.  Thus the zero-curvature equation for some
Lie algebra $g$ along with the covariant constancy conditions for a
set of differential forms taking values in some module over $g$ give
the simplest example of a free differential algebra. Note that this
case is not necessarily dynamically trivial and can be used to
describe some not empty free field dynamics. Here the zero curvature
equation for gauge fields of $g$ describes some $g$-invariant
background geometry while the covariant constancy conditions can
describe free fields propagating in this geometry provided that the
representation $t$ is infinite-dimensional.  For example, the free
massless equations (\ref{meq}) $ D_0 |\Phi(x) \rangle =0$ have
unfolded form. The connection $\go_0$ in (\ref{meq}) trivially
satisfies the zero curvature equation $ R =d\go_0 +\go_0\wedge \go_0 =
0 $ because it is $x$-independent and involves only mutually commuting
oscillators $a$. In fact, instead of fixing $\go_0$ in the particular
form (\ref{meq}) it is enough to say that the frame part of $\go_0$ is
nondegenerate and the condition (\ref{R0}) is satisfied, that brings
us to the formulation of the dynamics of massless fields in an arbitrary
conformally flat background.

Equations~(\ref{eq}) express exterior differentials of all fields $W^a
(x)$ via values of the fields themselves. As a result, to reconstruct
$W^a (x)$ everywhere by virtue of (\ref{eq}) modulo gauge ambiguity
(\ref{delw}), it is enough to fix values of $W^a (x_0)$ at any given
point\footnote{Of course, this is only true in a topologically trivial
  situation with trivial cohomology of the de Rahm differential $d$.
  Otherwise further restrictions on $W^a (x_0)$ can occur.}  $x_0$.
For this to be possible, the set of fields $W^a (x)$ must be rich
enough to describe all on-mass-shell nontrivial combinations of the
space-time derivatives of dynamical fields under consideration. {}From
the Poincare' lemma it follows that nontrivial degrees of freedom are
contained in the 0-forms because all degrees of freedom in $p-$forms with
$p>0$ are pure gauge, i.e., are fixed in terms of 0-forms modulo gauge
ambiguity.  Identification of field strengths with the 0-forms $C$
occurs in terms of some deformation of the zero-curvature equation of
the type~(\ref{R0}) to
\be
\label{def}
\R \equiv d\go +\go \wedge \go +\go\wedge \go C +\ldots =0
\ee
which, in fact, is the starting point towards the full nonlinear
dynamics in the form (\ref{eq})~\cite{unf}.  In HS gauge theories,
0-forms are described by the generating functions $C(a,b|x)$ analogous
to those of gauge potentials (\ref{expgo}) but taking values in a
certain ``twisted adjoint'' representation of the HS algebra. In the
conformal model discussed in this talk, the set of ``dynamical
fields'' $C(0,\bar{b}|x)$ and $C({b},0 |x)$ also corresponds to all
on-mass-shell nontrivial combinations of the gauge invariant fields
strengths of massless fields.

Note that nontrivial deformations (\ref{def})  sometimes
result from some zero-curvature and covariant constancy conditions
in a larger system
\be
\label{covc}
d\W = \W\wedge \W,\qquad D\C = 0
\ee
provided that the 0-forms $\C$ satisfy some nonlinear
constraints
\be
\label{con}
\varphi (\C)=0,
\ee
being invariant in the sense that $D\varphi (\C)=0$ as a consequence
of (\ref{covc}) (for more details and examples of HS gauge theories
formulated this way see~\cite{4d,Gol}).  In this case the nontrivial
dynamical content of the equations is hidden in the constraints
(\ref{con}).  The role of Eqs.~(\ref{covc}), which can be integrated
in the pure gauge form (at least locally), is to map the content of
the constraints (\ref{con}) to the geometric framework associated with
the original coordinates $x$. Such a formulation makes it trivial to
extend the dynamical equations to larger (super)spaces without
changing its dynamical content by simply adding extra coordinates $y$
and extending the differential forms to the extra dimensions
\be
\label{extra}
dx^n \W_n (x) \to dx^n \W_n(x,y) + dy^{n^\prime} \W_{n^\prime}(x,y) .
\ee
The dynamical content of the system (\ref{covc}), (\ref{con})
is still  encoded by the constraints (\ref{con}) on the zero forms
$\C$ at any point, say $\C(0,0)$, of a larger space-time. This trick
was suggested  in~\cite{BHS} where it was applied to the analysis
of the HS dynamics in the $sp(8)$ invariant space-time as
we discuss below. More recently, it was applied
in~\cite{SSS} to derive the superspace form of the $4d$ HS equations
of~\cite{4d}. As discussed in~\cite{BHS} there is a great
variety of extensions of the HS dynamics to different space-times.
The invariant content to be intact is the constraints (\ref{con})
\cite{4d}.

\section{Symmetries}
\label{Symmetries}
A nice feature of the unfolded formulation is that it makes symmetries
(\ref{delw}) manifest.  Suppose that some vacuum solution $W_0(x)$ of
(\ref{eq}) is fixed.  This restricts the gauge symmetry parameters by
the condition $\delta W_0(x)=0$ equivalent to
\be
\label{glob}
 d \e^a -\e^b
\frac{\delta F^a }{\delta W^b}\Big |_{W=W_0}=0.
\ee
If the 0-form part of the l.h.s. of (\ref{glob}),
$
\e^b
\frac{\delta F^a }{\delta W^b}\Big |_{W=C_0}
$,
is nonzero for a chosen solution, this imposes some restrictions
on the parameter $\e^a$. In many cases however $F^a (W)$ is such that
$\frac{\delta F^a }{\delta W^b}\Big |_{W=0} =0$.
Choosing a solution with  $C_0 =0$ one finds that for this case
the left hand side of Eq.~(\ref{glob}) is only
nontrivial  in the sector of $p$-forms with $p>0$ which is
a consistent differential condition because
 $\R (W_0)=0$. Therefore, it can be
integrated to fix $\e^a (x)$ in terms of $\e^a (x_0)$
 at any given point $x_0$ modulo
second rank ``gauge symmetries'' for gauge parameters
$
 \delta \e^a =d \xi^a +\xi^b
\frac{\delta F^a }{\delta W^b}.
$
{}From the Poincar\'e lemma it follows that nontrivial leftover
symmetries are parametrized by the 0-form parameters among
$\e^a$. These are reconstructed uniquely in terms of their
values $\e^a(x_0)$, which are
arbitrary parameters of the global symmetries of the system.
As a result, there are  as many global symmetries as
0-forms among $\e^a(x)$ at any fixed $x$,
i.e., as 1-forms among~$W^a$.

Applying this machinery to the HS system in
Sec.~\ref{$4d$ conformal free fields} we observe that
the Fock space forms a module over the
$4d$ conformal algebra $su(2,2)$ realized by the generators
\be
P_{\ga\dgb} = a_\ga \bar{a}_\dgb,\qquad
K^{\ga\dgb} = b^\ga \bar{b}^\dgb,\qquad
\ee
\be
L_\ga{}^\gb = a_\ga b^\gb -\half \delta_\ga^\gb a_\gga b^\gga,\quad
\bar{L}_\dga{}^\dgb = \bar{a}_\dga \bar{b}^\dgb
-\half \delta_\dga^\dgb \bar{a}_{\dot{\gga}} \bar{b}^{\dot{\gga}},\quad
D=\{a_\ga, b^\ga\} +\{\bar{a}_\dga, \bar{b}^\dga \}.
\ee
The 1-form $\go_0$ in (\ref{meq}) is the flat connection
of the conformal algebra. According to the general
analysis, this proves that Eq.~(\ref{meq}) is
conformally invariant. Since the conformal algebra $su(2,2)$
belongs to the symplectic algebra $sp(8)$
(\ref{sp}) which, in its turn,
belongs to the infinite-dimensional HS algebra of all polynomials
of the oscillators which  acts on the Fock module,
from our analysis it follows that the set of
equations for massless fields of all spins contained in (\ref{meq})
has global symmetries which form these algebras. Thus the
infinite-dimensional algebra of polynomials of oscillators of $a_A$
and $b^B$ is shown~\cite{BHS} to form a
symmetry of the system of equations of all massless fields that
extends the usual conformal symmetry. This confirms the conjecture
of Fradkin and Linetsky~\cite{FLA} that (an appropriate reduction
of) this algebra can be used as $4d$ conformal HS
algebra (these authors studied  a different
$4d$ conformal model being the HS extension of the
nonunitary $C^2$ conformal gravity.)
Algebras of this type were used in~\cite{SS1,5d} as $5d$ HS algebras.

The unfolding machinery provides
efficient tools for elucidating explicit form of
the transformation laws. For example, for the conformal HS system
in Sec.~\ref{$4d$ conformal free fields},
the zero curvature connection
admits the pure gauge representation
\be
\go_0 = g^{-1} dg ,\qquad g = \exp{-x^{\ga\dgb} a_\ga \bar{a}_\dgb}.
\ee
Then, the global symmetry parameter that leaves $\go_0$
invariant is
\be
\label{1}
\gvep (a,\bar{a},b,\bar{b}|x) = g^{-1}(a,\bar{a},b,\bar{b}|x)
 \gvep_0 (a,\bar{a},b,\bar{b})g(a,\bar{a},b,\bar{b}|x),
\ee
where $\gvep_0 (a,\bar{a},b,\bar{b})$ is an arbitrary $x$-independent
element of the HS algebra of oscillators which parametrizes the
HS global symmetries. The transformation law for massless fields
is
\be
\label{2}
\delta |\Phi \rangle = \gvep \delta |\Phi \rangle.  \ee The parameters
$\gvep_0 (a,\bar{a},b,\bar{b})$ bilinear in oscillators describe the
$sp(8)$ symmetry.  Those of higher orders describe HS symmetries.
Note that the parameters $\gvep_0 (a,\bar{a},b,\bar{b})$ being
order-$n$ polynomials in the oscillators give rise to some local
transformation laws with at most order $[\half n]$ space-time
derivatives.  Formulas (\ref{1}) and (\ref{2}) allow one easily to
derive explicit form of the respective transformation laws~\cite{BHS}.

\section{ $\sigma_-$ cohomology}
\label{cohomology}

A useful tool of the unfolding machinery is the identification
of the dynamical content of unfolded equations with
certain cohomology groups. Consider a covariant constancy
equation of the form
\be \label{geneq}
(\D +\sigma_- +\sigma_+)C (X)= 0 ,
\ee
where  $\D$ and $\sigma_\pm$ satisfy the relations
\be
\label{ds} (\sigma_\pm)^2 =0,\qquad \D^2 +\{\gs_-, \gs_+\} =0,
\qquad \{ \D, \sigma_\pm \} =0,
\ee
which guarantee that the connection is flat.  It is demanded that only
the operator $\D$, that contains the de Rahm differential, acts
nontrivially (differentiates) on the space-time coordinates while
$\sigma_\pm$ act in the fiber $V$ in which $C(X)$ takes values. It is
also assumed that there exists a grading operator $G$ diagonalizable
in $V$ such that its spectrum in $V$ is bounded from below and
\be \label{grad}
[G, \D ] = 0,\qquad [G
,\sigma_\pm ] = \pm \sigma_\pm.
\ee
For example, the conformal equations (\ref{oldeq}) have this form with
$\D = d$, $\sigma_+ =0$,
$\gs_- = -
dx^{\ga\dgb}\frac{\p^2}{\p b^\ga \p \bar{b}^\dgb}$ and
$G= \half \Big (b^\ga \frac{\p}{\p b^\ga} +
\bar{b}^\dga \frac{\p}{\p \bar{b}^\dga}\Big)$.
 In fact, it is a typical situation with
$V$ realized as a space of polynomials of some auxiliary
variables, $\gs_\pm$
being some differential operators acting on these
auxiliary variables and $G$ counting
a (nonnegative) polynomial degree that guarantees
that the spectrum of $G$ is bounded from below.

The following useful facts are true.
(i)~Dynamical fields contained in the 0-forms $C$ take values in the
  cohomology group $H^0 (\sigma_-)$. In other words, dynamical fields
  in $X$-space are those satisfying $\sigma_- (C(X))= 0$. This is
  because all fields in $V/H^0 (\sigma_-)$ (i.e., such that $\sigma_-
  (C(X))\neq 0$) are auxiliary being expressed via the space-time
  derivatives ($\D$) of the dynamical fields by virtue of
  Eqs.~(\ref{geneq}). Note that $H^0 (\sigma_-)$ is always nonzero
  because it at least contains the nontrivial subspace of $V$ of
  minimal grade.

(ii)~There are as many differential conditions on the dynamical
  0-forms $C$ in Eq.~(\ref{geneq}) as elements of the cohomology group
  $H^1 (\sigma_-)$.  If grade $k$ elements of $H^1 (\sigma_-)$ impose
  equations on grade $l$ dynamical fields, an order of differential
  equations is $k+1-l$.  All other equations in (\ref{geneq}) are
  either constraints, which express auxiliary fields via derivatives
  of the dynamical ones, or consequences of the dynamical equations.
  When $H^1 (\sigma_-)$ is zero (\ref{geneq}) is equivalent to some
  (usually infinite) set of constraints which express all fields
  contained in $C(X)$ via derivatives of the dynamical fields (which
  however have no dynamics in that case being restricted by no
  differential equations).


The proof is elementary~\cite{BHS}.
Indeed, consider the decomposition of the space of 0-forms $C$
into the direct sum of eigenspaces of $G$. Let a
field having definite eigenvalue $s_k$ of $G$ be
denoted $C_k$, $k= 0,1,2 \ldots$. Suppose that the dynamical
content of the equations (\ref{geneq}) with the eigenvalues
$s_k$ with $k \leq k_q $ is found. Applying the operator $\D $
to the left hand side
of  the equations (\ref{geneq}) at  $k \leq k_q $ we obtain
taking into account (\ref{ds})
that
\be
\label{sdc}
\sigma_- (\D ( C_{k_q+1} )) =0\,.
\ee
Therefore $\D ( C_{k_q +1} )$ is
$\sigma_-$ closed. If the group $H^1 (\sigma_- )$ is trivial
in the grade $k_q+1$
sector,  any solution of (\ref{sdc}) can be written in
the form
$\D ( C_{k_q+1} ) = \sigma_- \tilde{C}_{k_q +2}$
for some  field $\tilde{C}_{k_q +2}$. This,
in turn, is equivalent to the statement that one can adjust
$C_{k_q +2}$ in such a way
that $\tilde{C}_{k_q +2} =0$ or, equivalently, that the part of the
equation (\ref{geneq}) of the grade $k_q+1$ is some constraint
that expresses $C_{k_q +2}$ in terms of the derivatives of $C_{k_q +1}$
(here the assumption is used that
the operator $\sigma_-$ is algebraic
in the space-time sense, i.e. it does not contain space-time derivatives.)
If $H^1 (\sigma_- )$ is nontrivial, this means that the equation
(\ref{geneq}) sends the corresponding cohomology class to zero and,
therefore, not only
expresses the field $C_{k_q+2}$ in terms of  derivatives of
$C_{k_q+1}$ but also imposes some additional
differential conditions on $C_{k_q+1}$.
All other equations in (\ref{geneq})  either imply constraints
which express auxiliary fields via derivatives of the dynamical ones
or are consequences of the dynamical equations.

For example, for the system of massless fields (\ref{oldeq}) with
$\sigma_- = -\frac{ \p^2}{\p b^\ga \p\bar{b}^\dgb}$, $H^0 (\sigma_-)$
consists of analytic and antianalytic functions giving rise to the
dynamical massless fields $C(b,0|x)$ and $C(0,\bar{b}|x)$. The
dynamical equations (\ref{masseq}) along with the Klein--Gordon
equation are associated with $H^1(\sigma_-)$.

The following comments are now in order.

If $C(X)$ are $p$-forms, dynamical fields and
nontrivial dynamical equations are associated
with $H^{p}$ and $H^{p+1} (\gs_-)$, respectively.

Suppose that the symmetry algebra $g$
admits a triple $Z$ graded structure
$g = g_0 \oplus g_- \oplus g_+$ with Abelian subalgebras $g_\pm$. Let
$\gs_- = dX^A P_A$ where $P_A$ is some basis in $g_-$.
The dynamical fields satisfying
$\gs_- (C)=0$ then identify with the
primary fields in $C(X)$ satisfying $P_A C(X)=0$.\footnote{Note
the roles of ``translations" $P_A$
and ``special conformal" generators $K^A$ of $g_+$ acting in the
fiber is exchanged  in the unfolded formulation
compared to the standard induced representation
approach~\cite{ind} in which primaries are defined directly in
 the base manifold.} In other words, $H^0 (\gs_-)$
consists of singular vectors (i.e., vacua) of various submodules
over $g$ in $V$, which is consistent since any invariant submodule
of $V$ gives rise to a subsystem in (\ref{geneq}).
$H^0 (\gs_-)$
forms some representation of $g_0$. In most interesting physical
applications $H^0 (\gs_-)$ decomposes into (may be infinite)
direct sum of finite-dimensional representations of $g_0$.
This corresponds to a model with dynamical fields carrying
finite dimensional representations of $g_0$. Note that
for the case of usual conformal algebra, $g_0$ consists of
the Lorentz algebra plus dilatations, i.e., dynamical fields are
some Lorentz tensors carrying definite conformal weights.

Suppose that $V_{1,2}=V_1\otimes V_2$. Let $ C_1(X)\in V_1$
and $ C_2 (X)\in V_2$ solve (\ref{geneq}) with some operators
$ \gs^1_{-}$ and $ \gs^2_{-}$. Then
\be
\label{CCC}
C_{1,2} (X) = C_1 (X)\otimes C_2 (X)
\ee
solves (\ref{geneq}) with
$
\gs^{1,2}_{-} = \gs^1_- \otimes Id +Id\otimes \gs^2_-.
$
 If $V_1$ and $V_2$ form modules
over some symmetry algebra $g$, the same is true for $V_{1,2}$.
As a result, the unfolded formulation of the dynamical equations
equips the variety of solutions of
$g$-quasiinvariant partial differential
equations with the natural associative product structure isomorphic
to the tensor algebra of (semiinfinite)
modules over $g$. Note that this associative
structure differs from that of the ring of solutions
of first order differential equations. It maps solutions of
some set of (not necessarily first order) $g$-quasiinvariant
partial differential equations to solutions of some other
$g$-quasiinvariant equations. In fact, this property is deeply
related to the general $AdS/CFT$ philosophy~\cite{GV}.

\section{Field equations in $Sp(2M)$ invariant space-time}

{}According to the general argument in Sec.~\ref{Symmetries}
the set of massless equations of all spins
(\ref{meq})
admits $sp(8)$ symmetry realized by local field transformations.
Definite spin field equations are singled out by
the condition (\ref{ncond}). The operator $N$ in (\ref{ncond})
does not commute to the $sp(8)$ generators,
breaking $sp(8)$ down to the conformal algebra $su(2,2)$ which
is the centralizer of $N$ in $sp(8)$. The
$su(2,2)$ acts on every spin individually. The
generators in $sp(8)/su(2,2)$ mix different spins. There are
two irreducible subspaces of $sp(8)$ in the set of massless fields
of all spins, those containing  bosons or fermions (they form
an irreducible representation of $osp(1,8)$). Note that
 the Fock representation of $sp(8)$ used in our
construction is the nonunitary dual of the unitary Fock
(i.e., singleton) representation of $sp(8)$. As explained in
\cite{ShV,BHS} this duality has a form of some Bogolyubov transform.
That the unitary Fock representation of $sp(8)$ reduces to
the collection of massless
representations of all spins of $su(2,2)$ was discussed, e.g., in
\cite{Fr1,BLS}.

Our goal is to reformulate $4d$ dynamics of massless fields of all
spins in an equivalent but manifestly $sp(8)$ invariant  form.
To this end we have to use a $sp(8)$ invariant space-time.
As a minimum this
will give us a technical device analogous to superspace
in the context of supersymmetric theories.
Very likely, however, this generalization may have
much deeper effect on our understanding of the space-time
geometry. Since it is designed
for the description of just those sets of HS massless fields that
appear in consistent nonlinear HS theories, the proposed
formulation has a good chance to be related to a symmetric
phase of a theory of fundamental interactions.

Recall that the compactified Minkowski space-time is
the coset space
$SO(d,2)/P$ where $P$ is the parabolic subgroup of $SO(d,2)$
generated by special conformal, Lorentz and dilatation  generators.
Usual Minkowski space is the maximal cell of $SO(d,2)/P$. Let us
proceed analogously with the symplectic group $Sp(M)$ choosing the
parabolic subgroup of $Sp(M)$ generated by the ``special conformal''
generators $K_{AB}$ and Lorentz + dilatation
generators $L_A{}^B$ in (\ref{sp}). The dimension of the resulting
compactified generalized space-time $\M_M = Sp(M)/P$ equals to
the number of translation generators in (\ref{sp}), i.e., $dim(\M_M)=
\half M(M+1)$. Local coordinates are real symmetric
matrices $X^{AB}=X^{BA}$. This space was discussed by many authors
in different contexts~\cite{gensp,Fr1,Ced,Gun,GGHT}. It allows
the action of $sp(2M)$  realized by the vector fields
 \be
P_{AB}=\frac{\partial}{\partial X^{AB}},\qquad
L_{A}{}^{B}=X^{BC}\frac{\partial}{\partial
X^{AC}},\qquad
K^{AB}=X^{AC}X^{BD}\frac{\partial}{\partial X^{CD}}.
\ee
In this talk we will not distinguish between the
compactified space $\M_M$ and its maximal cell having trivial
topology of $R^{\half M(M+1)}$.

Let us now extend Eq.~(\ref{meq}) to $\M_4$ with the
help of the unfolded formalism machinery.
This is achieved by means of the trick explained at the end of
Sec.~\ref{Unfolded dynamics} via replacing  (\ref{meq}) with
\be
\label{meqgen}
D_0 |\Phi (X) \rangle \equiv d |\Phi (X) \rangle +\gs_-
|\Phi (X) \rangle =0,
\quad \gs_- = - dX^{AB} \f{\p^2}{\p b^A \p b^B}.
\ee
($d = dX^{AB}\f{\p }{\p X^{AB}}$).
Decomposing  $X^{AB}$ as
$x^{\ga\dga}, x^{\ga\gb}, \bar{x}^{\dga\dgb}$ with Hermitian
coordinates $x^{\ga\dga}$ of the usual Minkowski space-time
and additional mutually conjugated six coordinates $x^{\ga\gb}$ and
$\bar{x}^{\dga\dgb}$
we find that this equation contains the original
$4d$ equation (\ref{oldeq})  along with the equations
$
\frac{\p}{\p x^{\ga\gb}} C(b|X) - \frac{\p^2}{\p b^\ga \p {b}^\gb}
C(b|X) =0\,
$
and their conjugates. Clearly these new equations just reconstruct
the dependence on the additional six coordinates $x^{\ga\gb}$
and $\bar{x}^{\dga\dgb}$ in terms of values of the original
fields in the $4d$ Minknowski space-time.

Now we are in a position to  analyse Eq.~(\ref{meqgen})
directly in the ten dimensional space-time $\M_4$.
The analysis of its dynamical content
is elementary in terms of $\sigma_-$ cohomology.
Let us consider the case of arbitrary $M$.
Obviously, $H^0 (\gs_-)$ is spanned by at most linear polynomials
in $b$, i.e., dynamical fields are
\be
C^{dyn} = b(X) + b^A f_A (X).
\ee
Here the scalar field $b(X)$ and spinor field $f_A (X)$
in $\M_M$ describe, respectively, all massless boson and
fermion fields in the  $4d$ Minkowski space-time in the case of
$M=4$. It is not hard to see (explicit proof is given in~\cite{GV})
that $H^1 (\sigma_-) $ is spanned by the elements of
the form
$
\ F_{C D,B} b^B dX^{C D},
$ and $B_{C D,B A} b^B b^A dX^{C D},$
where   $F_{C D,B}$ and $B_{C D,B A}$
are arbitrary tensors having symmetry properties of the
Young tableaux
\quad
\begin{picture}(13,12)(0,0)
{\linethickness{.250mm}
\put(00,00){\line(1,0){05}}
\put(00,05){\line(1,0){10}}
\put(00,10){\line(1,0){10}}
\put(00,00){\line(0,1){10}}
\put(05,00){\line(0,1){10}}
\put(10,05){\line(0,1){05}}
}
\end{picture}
and
\quad
\begin{picture}(13,12)(0,0)
{\linethickness{.250mm}
\put(00,10){\line(1,0){10}}
\put(00,05){\line(1,0){10}}
\put(00,00){\line(1,0){10}}
\put(00,00){\line(0,1){10}}
\put(05,00){\line(0,1){10}}
\put(10,00){\line(0,1){10}}
}
\end{picture}
, respectively.
The nontrivial equations on the dynamical fields, which
have these symmetry properties,
are \cite{BHS}
 \be
\label{oscal} \Big (
\f{\p^2}{\p X^{AB} \p X^{CD}} - \f{\p^2}{\p X^{AC} \p
X^{BD}}\Big) b(X) =0
\ee
for the boson field $b(X)$ and
\be
\label{ofer} \f{\p}{\p X^{AB}} f_C(X) -
\f{\p}{\p X^{AC}} f_B(X)      =0
\ee
for the fermion field $f_B(X)$.
These are analogues of the Klein--Gordon and Dirac equations.
(For  $M=2$ they  coincide with the $3d$ massless equations.)
Note that every solution of (\ref{ofer}) satisfies (\ref{oscal}).

As Eqs.~(\ref{oscal}) and (\ref{ofer})
with $M=4$ are equivalent to the
original $4d$ equations they are expected to respect nice
features of the original $4d$ system, such as unitarity, locality,
microcausality etc. It is instructive to reanalyse these
properties  starting directly from
Eqs.~(\ref{oscal}) and (\ref{ofer}). Once these are
linear equations one can analyze them via Fourier transform.
For a particular harmonic
\be
\label{expb}
b(X) = b_0 \exp i k_{AB}X^{AB}\,
\ee
 (\ref{oscal}) requires
\be
\label{kk}
k_{AB}k_{CD}=k_{AC}k_{BD}.
\ee
This is solved by the twistor ansatz
\be
\label{ktw}
k_{AB} =k \xi_A \xi_B
\ee
with an arbitrary commuting real ``svector'' $\xi_A$ and a factor $k$.
For the proof it is enough to diagonalize the
symmetric matrix $k_{AB}$ by a $SL_M$ transformation to see that
the product of any two different eigenvalues is zero by  (\ref{kk})
at $A \neq B$ and, therefore,
any nonzero matrix $k_{AB}$ satisfying (\ref{kk}) has rank 1.
 Modulo rescalings of $k$ and $\xi_A$,
there are two essentially different options in (\ref{ktw}), with
$k=1$ or $k=-1$. These correspond to the positive and negative
frequency solutions, respectively. The analysis of fermions is
analogous. As a result the generic solutions of (\ref{oscal}) and
(\ref{ofer}) have the form
\be
\label{bfo}
b (X) =\f{1}{\pi^{\f{M}{2}}}
\int d^M\xi\, \Big (b^+ (\xi) \exp i  \xi_{A}\xi_{B} X^{AB}
+b^- (\xi) \exp -i  \xi_{A}\xi_{B}X^{AB} \Big) ,
\ee
\be
\label{ffo}
f_C (X) =\f{1}{\pi^{\f{M}{2}}}
\int d^M\xi\, \xi_C
\Big (f^+ (\xi) \exp i  \xi_{A}\xi_{B} X^{AB}
+ f^- (\xi) \exp -i  \xi_{A}\xi_{B}X^{AB} \Big) .
\ee
Both for the scalar $b(X)$ and svector $f_A (X)$,
the space of solutions is parametrized by two
functions of $M$ variables $\xi_A$. Because odd functions
$b^\pm (\xi)$ and even functions
$f^\pm (\xi)$ do not contribute to (\ref{bfo})
and (\ref{ffo}), respectively, we require \be
\label{par} b^\pm (\xi)=b^\pm (-\xi) ,\qquad
f^\pm (\xi)=-f^\pm (-\xi).  \ee
The integration in (\ref{bfo}) and (\ref{ffo}) is thus carried out over
$R^M / Z_2$. The origin of coordinates $\xi_A =0$ is
invariant under the $Z_2$
reflection $\xi_A\to -\xi_A$, being a singular
point of the conical orbifold $R^M / Z_2$. Note that the whole
framework is supersymmetric as is manifested by the fact
that
bosons and fermions have equal numbers
of degrees of freedom  for any $M$.

As shown in~\cite{Mar}, the form of the general solution (\ref{ffo}),
(\ref{bfo}) allows  manifestly $sp(2M)$ covariant quantization.
The commutation relations for quantized fields
equivalent to those in $3d$ and $4d$
Minkowski space-time for $M=2$ and $M=4$, respectively,
are as simple as
\be
\label{qpr}
[b^\pm (\xi_1), b^\pm (\xi_2)] =0,\qquad
[b^- (\xi_1), b^+ (\xi_2)] =
\half (\delta (\xi_1 - \xi_2)+\delta (\xi_1 + \xi_2)),
\ee
\be
\label{qprf}
[f^\pm (\xi_1), f^\pm (\xi_2)]_+ =0,\qquad
[f^- (\xi_1), f^+ (\xi_2)]_+ =
\half (\delta (\xi_1 - \xi_2)- \delta (\xi_1 + \xi_2)),
\ee
while the Green function is $det|X_1 - X_2|^{-\half}$.

\section{Time}

Equations~(\ref{oscal}) and (\ref{ofer}) describe propagation along
the generalized light-like directions $ \Delta X^{AB} = \eta^A
\eta^B, $ where $\eta^A$ is a contravariant ``svector''.  The sign
choice here fixes a time arrow.  Suppose that a light-like signal
emitted from some point $X_0^{AB}$ of the generalized space-time
reaches some other point $X_1^{AB}=X_0^{AB} + \eta^A \eta^B$ switching
on a new process that emits a signal in a different light-like
direction\footnote{The assumption that a process can be switched on
  locally may or may not be true for particular dynamical equations.
  In fact, as discussed below, Eqs.~(\ref{oscal}) and~(\ref{ofer}) do not
  admit full localization in $\M_M$.  This does not affect our
  conclusions on the causal structure of the generalized space-time,
  however.}.  Provided this happens several times, any point \be
\label{del}
\Delta X^{AB} =\sum_{i=0}^{M} \eta_i^A \eta_i^B,
\ee
can be reached
where $\eta_i^A$ is a complete set of contravariant svectors.
Formula (\ref{del}) describes  generic
positive semi-definite symmetric matrix $\Delta X^{AB}$.

We  see that the relativistic geometry that follows
from Eq.~(\ref{oscal}) identifies the future
cone $\C_{X_0}^+$ of a point $X_0$
with the set of matrices $X^{AB}$ such that
$\Delta X^{AB}=X^{AB}-X_0^{AB} $ is
positive semi-definite. Time-like vectors
are described by positive definite matrices
\be
\Delta X^{AB}\xi_A \xi_A >0\q \forall \xi_A. \nn
\ee
Light-like vectors identify with degenerate positive semi-definite
matrices.
We will distinguish between rank-$k$ light-like directions
described by matrices of rank $k$. The
concepts of time-like and rank-$k$ light-like vectors are invariant
under the generalized Lorentz group $SL_M$.

Equation (\ref{oscal}) describes propagation along the most degenerate
light-like directions of rank~$1$.  To work out a form of the
equations that describe propagation along less degenerate light-like
directions one can use the tensor product construction described at
the end of Sec.~\ref{cohomology}. As shown in~\cite{GV}, $n-$fold
tensor product of the Fock representation in Eq.~(\ref{meqgen}) gives
rise to the field equations which describe propagation along rank $n$
light-like directions in $\M_M$.

The past cone $\C_{X_0}^-$ is defined analogously as the set of
negative semi-definite matrices
\be
(X^{AB}-X_0^{AB}) \xi_A \xi_B \leq 0\q \forall \xi_A.
\nn
\ee
If $Y\in \C_{X}^+$ then $X \in \C_{Y}^-$ and
$2X-Y \in \C_{X}^-$.
$\C_{X}^+$ is the convex cone: $\forall X_1, X_2 \in
\C_{X}^+, \lambda,\mu \in R^+,
\lambda X_1 +\mu X_2 \in \C_{X}^+$. Note that the analysis
of future and past cones in $\M_M$ was given in~\cite{GGHT}
in a different context.

Let us now introduce a concept of space-like global Cauchy surface as
such a submanifold $\Sigma$ of some (generalized) space-time manifold
$\M$ that

\begin{enumerate}
\item[$(i)$] $\forall X_1, X_2 \in \Sigma $, $X_1 \notin \C^\pm_{X_2}$
  and $X_2 \notin \C^\pm_{X_1}$ for $X_1\neq X_2$.

\item[$(ii)$] any point of $\M$ belongs to either future or past cone
  of some point of $\Sigma$. No point $Y\in \M$ can belong to the
  future cone of some point $X_1 \in \Sigma$ and past cone of some
  other point $X_2 \in \Sigma$.
\end{enumerate}

In particular, the property $(i)$ implies that no pair of observers on
$\Sigma$ are allowed to exchange causal signals, i.e., a global Cauchy
surface is space-like.  Provided that $\M$ admits a fibration (or
rather foliation) into a set of space-like global Cauchy surfaces,
$\Sigma_t$ parametrized by some parameter(s) $t$, this defines the
concept of time(s).

Let $T^{AB}$ be some fixed  positive definite matrix.
The axioms $(i)$ and $(ii)$ are satisfied with the
space-like global Cauchy surfaces $\Sigma_t$ parametrized as
\be
\label{res}
X^{AB} \in \Sigma_t :\qquad X^{AB} = x^{AB}+tT^{AB},
\ee
where the space coordinates $x^{AB}$ are
arbitrary $T-$ traceless matrices
\be
\label{xtr}
x^{AB}T_{AB}=0,\qquad T_{AB}T^{BC} =\delta_A^C.
\ee
Indeed, the difference of any two matrices
of the form (\ref{res}) at fixed $t$ is traceless and
therefore it is neither positive definite nor negative definite.
As a result, any two points of $\Sigma_t$  at some fixed $t$
are separated by a space-like interval.
The rest of the axioms is a consequence of the trivial decomposition
(\ref{res}) of a matrix into the sum of its trace and traceless parts.

An important output of this analysis is that
$\M_M$ has just one evolution parameter
\be
\label{t}
t=\f{1}{M} X^{AB} T_{AB}.
\ee
The ambiguity in the choice of a positive definite
matrix $T^{AB}$ parameterizes the ambiguity in the choice of a
particular coordinate frame like in Einstein special
relativity: any two positive definite
matrices $T_{1,2}^{AB}$ with equal
determinants are related by some
generalized $SL_M$ Lorentz transformation. The dilatation
allows one to fix a scale of time in an arbitrary way.
As shown in~\cite{Mar}, $SL_M$
generalized Lorentz transformations exhibit a lot of similarity
with the usual Lorentz transformations in Minkowski space-time including
nontrivial limits on relative speeds of reference frames.

\section{Localization and Clifford algebra}
\label{Localization}

Using the ambiguity in $b^\pm (\xi) $ and $f^\pm (\xi)$
in the general solutions (\ref{qpr}) and (\ref{qprf})
it is possible to fix their dependence in at most
$M$ coordinates that, generically, is much less than
$dim(\M_M) = \half M(M+1) $. In particular, it is impossible
to localize solutions in more than $M$ coordinates.
This phenomenon is new
compared to the experience with the usual Minkowski space-time:
although fields live in $\half M(M+1) $ dimensional space-time
physical events occur in some
$M$-dimensional subspace. This happens because
the equations are overdetermined and contain constraints
which do not allow
to focus the waves in more than $M$
dimensions. (Analogously, the Gauss law does not allow electric field
to be fixed arbitrarily  on the initial Cauchy surface
for a given charge distribution.)

It is useful to introduce~\cite{Mar} the concept of local
Cauchy bundle $E$ as such  $M$-dimensional
surface in $\M_M$ (or a limiting
surface with some of radii shrinking to zero) that the solutions
(\ref{qpr}) and (\ref{qprf}) are fixed by their values on $E$.
Note that while the concept of global Cauchy surface is a
characteristics of $\M_M$ independent of  particular dynamics,
the concept of local Cauchy surface characterizes particular
dynamical equations in $\M_M$.
These two notions coincide in the Minkowski geometry
but turn out to be different in a more general framework.

Now one can address a question whether  there are some $d-1$
space-like coordinates
\be
x^n = \sigma^n_{AB}X^{AB}
\ee
such
that, using the ambiguity in the functions $b^\pm (\xi)$, it is
possible to built  solutions of the field equations
proportional to (derivatives of)
the delta function $\delta^{d-1}(x-x_0)$
at any point $x_0 \in R^{d-1}$. For this to be possible, the
dual momenta $k_n = \Gamma_n^{AB}\xi_A\xi_B$ have to provide a
map of the orbifold $R^M/Z_2$ onto $R^{d-1}$ with the only singularity
at $\xi_A =0$. Then, by changing integration variables from $\xi_A$
to $k_n$ plus, may be,  some other variables in case
$d-1 <M$, one can get delta functions $\delta^{d-1}(x-x_0)$
by integrating over $k_n$.

Remarkably, this can be achieved if the matrices
$\gamma^n{}_{A}{}^B = \gs^n_{AC}T^{CB}$,
where $T^{CB}$ is the time axis matrix, satisfy Clifford
relations
\be
\label{clif}
\{\gamma^n, \gamma^m \} = 2\eta^{nm}.
\ee
(This  is possible when the index $A$ describes
a collection of some $q$ spinors with $M=q 2^p$.) Indeed,
an immediate consequence of
(\ref{clif}) is  that the matrices $\gs^n_{AB}$ are traceless
$
\gs^n_{AB} T^{AB} =0\,
$
whenever $d\geq 3$, thus belonging to the global Cauchy surface.
Another important property is that the momenta
\be
\label{mom}
k_n (\xi) = \sigma_n^{AB} \xi_A\xi_B\,
\ee
map the cone $R^M /Z_2$ onto $R^{d-1}$, i.e., varying real twistor parameters
$\xi_A$ it is possible to get arbitrary values of $k_n (\xi) $. This
results from the invariance of the construction under the space rotations
$SO(d-1)$ generated by
$
M_{nm}= \f{1}{4}[\gamma_n, \gamma_m ].
$
By a space rotation one aligns a vector $k_n (\xi)$ along any
direction and then normalizes it arbitrarily by a rescaling of $\xi_A$.
That momenta $k_n (\xi)$
span $R^{d-1}$ allows for localization of the fields in $d-1$ space
$x^n$ coordinates dual to $k_n$, i.e., by means of integration over
$k_n$, one can reach the delta-functional initial data
$\delta^{d-1} (x^n - x^n_0)$ localized at any point of the
physical space $R^{d-1}$~\cite{Mar}.

Because the square of any linear combination of
 $\gamma$ matrices is proportional to the unit matrix, for any vector
$a^n$ there exists such a basis in the space of $\xi_A$ that
\be
\label{keyrep}
T^{AB}= \delta^{AB},\qquad a^n \sigma_n^{AB} =
\sqrt{a^2} Y^{AB},\qquad a^2 =a^n a^m \eta_{nm},
\ee
where
\be
Y= \left(\begin{array}{cc}
                   I    &   0   \\
                   0    &  -I
            \end{array}
    \right)  \nn
\ee
with all four blocks being $\f{M}{2} \times \f{M}{2} $ matrices
($M$ is assumed to be even). As a result,
(nondegenerate) space-time matrix coordinates of the form
\be
\label{tX}
\chi^{AB} = t T^{AB} + x^n \sigma_n^{AB}
\ee
can only have three values of the inertia index
$
I_\chi = M$ for $ t>\sqrt{x^2},
$
$
I_\chi = - M$ for $ t< - \sqrt{x^2}
$
and
$
I_\chi = 0 $ for $ t^2 < {x^2},
$
where
$
x^2 =x^n x^m \eta_{nm} .
$
This corresponds to the standard space-time picture with the
future cone ($I_\chi >0$), past cone ($I_\chi <0$) and the space-like
region ($I_\chi =0$). This property of space Clifford coordinates
implies microcausality in terms of the coordinates $x^n$ of the
usual space-time~\cite{Mar}.
(Note that the condition that $I_\chi$ can take
in the linear space of matrices of the form (\ref{tX})
only maximal value $M$ (future), minimal value $-M$
(past) or some fixed  intermediate value
$I_\chi =I_{space}\neq \pm M$ for all values of $t$ and $x^n$
can be taken as an alternative definition leading to
 the $\gga$-matrix
relations~\cite{Mar}.)

A new comment we would like to make here is
that the Clifford  realization of
coordinates of local events provides a global
characterization of causality. Consider
two events characterized by local space-time coordinates (\ref{tX})
$\chi_1$ and $\chi_2$.  A
solution of the field equations localized
at $\chi=\chi_{1,2}$ is characterized by some function $c(X)$ of
\be
\label{chilam}
X^{AB}= \chi^{AB} +\Lambda^{AB}
\ee
with the dependence in $\chi$ localized at $\chi_{1,2}$
(i.e., being proportional to some derivatives of
$\delta (\chi - \chi_{1,2} )$) but not necessarily
localized in the rest
traceless coordinates $\Lambda^{AB}$ orthogonal to
$ \gs_{n\,AB} $
\be
\label{Lort}
\Lambda^{AB} \gs_{n\,AB} =0\,,\qquad
T^{AB} \gs_{n\,AB} =0\,.
\ee
The question is whether it is possible that $\chi^{AB}$ is
space-like
while $X^{AB}$ is time-like for some  $\Lambda^{AB}$.
Remarkably, the answer is negative (otherwise, causal
relationships beyond the usual local coordinate Minkowski
picture would be possible). Namely,
let $X_1 , X_2 \in \M_M$, be some
points where solutions corresponding to some two different
events are nonzero. If $X_1 \in \C_{X_2}^\pm$
the Clifford local coordinates
$\chi_{1,2}$ characterizing these solutions
are in the same causal relationship
(for any choice of Clifford coordinates). This fact guarantees
that the concept of causality has global meaning and
is fully characterized by the causality
of local events associated with the Clifford coordinates.

The proof is elementary. Let $X^{AB}\in \C,$ where $\C=\C_0^+$,
i.e. $X^{AB}$ is positive
definite. The dual cone $\C^*$ consists of all positive
definite matrices  $Y_{AB}$\footnote{Recall
that a dual cone $\C^*$ of $\C\subset V$, where $V$ is some
vector space, consists of such elements $a\in
V^*$ that the scalar product  $(a,b)>0$ $,\forall
b\in \C$. As is easy to see, $Y_{AB}
X^{AB}>0$ for any  positive definite matrices $X^{AB}$ and $Y_{AB}$,
and, if one of the matrices, say $Y_{AB}$, is not positive definite,
then there exists such a positive definite martrix  $X^{AB}$
that $Y_{AB} X^{AB} \leq 0$.}. The
two spaces
are identified once the matrix $T^{AB}$ is used to raise and
lower indices.
 Positive definite matrices $\chi^{AB}$ of the form (\ref{tX})
also form a selfdual  cone, which is the usual
future cone $t>0$, $t^2-x^2 >0$ in the Minkowski space-time.
It follows that ${\chi}^*_{AB}X^{AB} >0$ for any positive
definite ${\chi}^*_{AB}$. The property (\ref{Lort}) implies
that ${\chi}^*_{AB}\chi^{AB} >0$ for any ${\chi}^*_{AB}$.
Since the Minkowski cone is selfdual in Minkowski space-time,
 $\chi^{AB}$ belongs to the Minkowski cone and therefore
is positive definite.

These properties indicate that the Clifford algebra realization of
space coordinates has deep relationship with
the concept of locality and microcausality.
In other words, the generalized space-time $\M_M$ is visualized
via  Clifford algebras. Let us stress that there is no metric
tensor in the dynamical equations (\ref{oscal}) and (\ref{ofer}), i.e.,
the original $Sp(2M)$ invariant setup
is insensitive not only to the choice of the conformal factor
of the metric but does not contain the metric tensor at all.
The space metric
$\eta_{nm}$, which automatically turns out to be
positive definite as a consequence of (\ref{clif}),
appears in the theory  upon identification of an
appropriate Clifford algebra  at the stage of
defining  the concept of  local event. It is indeed physically
meaningful to define a physical length measure
(metric) simultaneously  with the  definition of a physical point.
Thus the formulation of HS dynamics in the generalized space-time
drives us to a point at which a
role of the metric tensor has to be reconsidered as not beeing given
{\it ad hoc} but resulting from the analysis of the concept of
locality
in a theory that has no metric tensor in its original formulation.
Let us stress that this in no way means that the the property of
independence of a coordinate choice (i.e., equivalence principle)
is lost. It is guaranteed by the unfolded form of the dynamical
equations (\ref{eq}) formulated in terms of differential forms.
The new message is that prior defining how to measure
a distance between events one has to specify what is a local physical
event.

 For the lowest values of $M=2,4,8$, the local Cauchy bundle $E$
 has the structure $E=\gs\times S$ with the base manifold
(local Cauchy surface)
$\gs=R^2, R^3, R^5$ as the physical space and
fiber compact manifolds $S= Z_2, S^1, SU(2) $,
respectively. This corresponds to
 $3d$, $4d$ and $6d$ Minkowski space-times,
while the fibers $S^1$ and $SU(2)$ give rise to some spin degrees
of freedom. In particular, for the case of $M=4$, modes of
 $S^1$ give rise to the infinite tower of spins
in $4d$ Minkowski space-time~\cite{Mar}.
Once some local Cauchy bundle $E= R^{d-1} \times S$ is chosen
to visualize  $\M_M$, the interpretation of different
subalgebras of the $Sp(2M)$ symmetry of the original
equations becomes different.
A  transformation which maps the Minkowski space-time
$\gs\times R^1$  to itself leaving the fibers intact
is some usual (Minkowski)
conformal transformation (conformal symmetry is a
subalgebra of $Sp(2M)$ as a consequence of
the Clifford realization of  space~\cite{Mar}).
Another type of symmetry
does not shift points of the Minkowski space-time acting only
on the coordinates of the fiber. For the case of $M=4$ this is
the $U(1)$ electro-magnetic duality transformation acting
on all spins, which thus acquires a purely geometric interpretation
in the generalized space-time framework. The
$Sp(2M)$ transformations which shift $E$  in
$\M_M$ look as nongeometric symmetries from the Minkowski
space-time perspective,  extending
$su(2,2)\oplus u(1)$ to $sp(8)$ which
 mixes fields of different spins.

\section{Conclusions}
The fundamental fact that infinite-dimensional HS symmetry algebras
are oscillator star-product algebras has a consequence that they
contain $sp(2M)$ as finite-dimensional subalgebras when there are $M$
pairs of oscillators.  For lower dimensions these are isomorphic to
the usual conformal or AdS space-time symmetry algebras.  For higher
dimensions, starting from the $4d$ conformal algebra (equivalently
$AdS_5$ algebra), space-time symmetry algebras become proper
subalgebras of the symplectic subalgebras of the HS algebras. In
particular, the $4d$ conformal algebra $su(2,2)$ is a subalgebra of
the $sp(8)$ that acts on the infinite towers of massless fields. The
idea that HS theories have to admit a description in a larger
manifestly $sp(8)$ invariant space-time is as natural as the idea to
describe supersymmetric theories in superspace. The relevant $Sp(2M)$
invariant space-time $\M_M$ turns out to be $\half M(M+1)$ dimensional
having $M\times M $ symmetric matrices $X^{AB}=X^{BA}$ as local
coordinates ($A,B= 1 \ldots M$). These coordinates contain usual
space-time coordinates $x^n$ along with coordinates $y$ analogous to
the central charge coordinates known to appear in the context of brane
dynamics and $M$-theory algebras (see, e.g.,
\cite{Curt,BLS,CAIP,GGHT,MKR} and references therein). The infinite
towers of $4d$ massless fields of all spins are described by one
scalar $b(X)$ for all bosons and one spinor $f_A (X)$ for all
fermions which satisfy, respectively, the second-order (\ref{oscal})
and first-order (\ref{ofer}) partial differential equations in $\M_4$
equivalent to all massless equations in the $4d$ Minkowski space-time.

Remarkably, these equations do not contain any metric tensor in
$\M_M$. This unusual feature is physically appropriate because the HS
equations in $\M_M$ fix solutions by their values on some
$M$-dimensional manifold called local Cauchy bundle $E$. One can say
that $E$ is the space of physical events.  On the top of that there is
one time evolution parameter associated with a time-like direction
along some positive definite matrix $T^{AB}$.  For $M=4$, $E= R^3
\times S^1$~\cite{Mar}. Modes on $S^1$ give rise to the infinite
towers of higher spins.  The electro-magnetic duality symmetry gets
geometric interpretation as a part of the $Sp(2M)$ symmetry which acts
on the fiber $S^1$.  Metric structure reappears in the theory at the
stage of identification of the space coordinates which describe local
events (i.e., $R^3$ for the $M=4$ case) in a quite remarkable way:
matrix coordinates which admit localization form Clifford algebra.
This introduces simultaneously the metric tensor and spinor structure
in the physical space-time of local events with the indices $A,B$
interpreted as (a collection of) spinor indices in the physical
space-time. The lesson is that having a mathematical space like $\M_M$
is not the same as to have a physical space-time where events
described by one or another equations can happen.  Different equations
in the same space can visualize it differently. Apart from the
simplest (rank 1) dynamical system in $\M_M$ discussed in this talk,
supported by $M$-dimensional local Cauchy surfaces, there exist higher
rank system supported by $rM$-dimensional local Cauchy surfaces
\cite{GV}.  In striking similarity with the brane picture in
superstring theory physical space-times visualized by different
equations coexist as ``branes'' imbedded into $\M_M$.

Reconsideration of such fundamental notions of physics as local event
and local distance (metric) suggested by the analysis of the HS gauge
theory may significantly affect the present day understanding of
space-time geometry and related physical theories including general
relativity and quantum mechanics.  The most exciting option is that
this may help to unify these two fundamental physical theories on a
new basis.

The main practical problem for the future is to develop a manifestly
$Sp(2M)$ symmetric nonlinear HS theory. In particular, this will
require to include gauge potentials into the formalism as the
equations discussed so far are formulated in terms of field strengths.

\paragraph{Acknowledgments.}
This research was supported in part by INTAS, Grant No.99-1-590 and by
the RFBR Grants No.02-02-17067 and 00-15-96566.

\end{document}